\documentclass[aip,jcp,floatfix,twocolumn,tightenlines,a4paper,10pt]{revtex4-1}
\usepackage[final]{graphicx}
\usepackage{amsmath}
\usepackage{amssymb}

\begin{document}

\title{High-resolution Fourier-transform XUV photoabsorption spectroscopy of $\boldsymbol{{}^{14}}$N$\boldsymbol{{}^{15}}$N}
\author{A.N. Heays}
\author{G.D. Dickenson}
\author{E.J. Salumbides} 
\affiliation{Institute for Lasers, Life and Biophotonics Amsterdam, VU University, De Boelelaan 1081, 1081 HV Amsterdam, The Netherlands}
\author{N. de Oliveira} 
\author{D. Joyeux} 
\author{L. Nahon}
\affiliation{Synchrotron Soleil, Orme des Merisiers, St. Aubin, BP 48, 91192 Gif sur Yvette Cedex, France}
\author{B.R. Lewis} 
\affiliation{Research School of Physics and Engineering, The Australian National University, Canberra, Australian Capital Territory 0200, Australia}
\author{W. Ubachs}
\affiliation{Institute for Lasers, Life and Biophotonics Amsterdam, VU University, De Boelelaan 1081, 1081 HV Amsterdam, The Netherlands}

\begin{abstract}
The first comprehensive high-resolution photoabsorption spectrum of ${}^{14}\rm{N}{}^{15}\rm{N}$ has been recorded using the Fourier-transform spectrometer attached to the Desirs beamline at the Soleil synchrotron.
Observations are made in the extreme ultraviolet (XUV) and span 100\,000--109\,000\,cm$^{-1}$ (100--91.7\,nm). 
The observed absorption lines have been assigned to 25 bands and reduced to a set of transition energies, $f$ values, and linewidths.
This analysis has verified the predictions of a theoretical model of N$_2$ that simulates its photoabsorption and photodissociation cross section by solution of an isotopomer independent formulation of the coupled-channel Schr\"odinger equation.
The mass dependence of predissociation linewidths and oscillator strengths is clearly evident and many local perturbations of transition energies, strengths, and widths within individual rotational series have been observed.
\end{abstract}

\maketitle

\section{Introduction}

Molecular nitrogen has been one of the most vigorously studied diatomic molecules but many details of its complex spectrum are still not understood, particularly with regard to its strongly and erratically perturbed absorption intensities and predissociation linewidths.
More detailed experimental knowledge of N$_2$ photoabsorption and photodissociation, and their isotopic dependence, would help clarify the quantum-mechanical picture of the molecule and is of particular application to a broad range of studies concerning atmospheric and astrophysical photo-chemistry.\cite{bishop_etal2007,liu_etal2009,lavvas_etal2011,stevens_etal2011}

On the experimental side, a variety of techniques have been employed to chart the dipole-allowed absorption spectrum of N$_2$ which begins at 100\,nm and extends to shorter wavelengths, thus restricting optical experiments to windowless techniques.
Electron-energy-loss spectroscopy has revealed many prominent features at low resolution\cite{zipf_mclaughlin1978,ajello_etal1989,khakoo_etal2008} which have been recorded with greater precision by means of classical spectroscopy in emission\cite{tilford_wilkinson1964,roncin_etal1998,roncin_etal1999} and in absorption.\cite{ogawa_tanaka1962,carroll_collins1969,carroll_yoshino1972}
The finest resolution has been achieved by using combinations of molecular beams and narrow-band tunable XUV laser sources.\cite{ubachs_etal1989,levelt_ubachs1992,sommavilla_etal2002}
Quantitative measurements of oscillator strength have only been made for absorption bands of the most abundant isotopomer, ${}^{14}\rm{N}_2$, and have been derived from grating-based spectrometry combined with synchrotron-generated XUV radiation.\cite{stark_etal2005,stark_etal2008,heays_etal2009} 
These follow earlier studies\cite{carter1972,gurtler_etal1977} which suffered from saturation effects due to insufficient resolution.
Measurements of predissociation rates have been obtained from linewidth studies using laser\cite{ubachs_etal1989,ubachs1997,ubachs_etal2000} and synchrotron\cite{stark_etal2005,stark_etal2008,heays_etal2009} sources, as well as pump-probe experiments with picosecond time resolution.\cite{ubachs_etal2001,sprengers_etal2004a,sprengers_etal2004b} 
The excitation of predissociative states from charge-exchanged molecular beams has also provided information about the excitation levels of atomic photofragments.\cite{walter_etal1993,buijsse_etal1996}

Much less is known about the less abundant N$_2$ isotopomers.
There have been several studies over the years of isotopically purified samples of $^{15}$N$_2$\cite{ogawa_etal1964,sprengers_etal2003,sprengers_ubachs2006} and only a few scattered observations of $^{14}$N$^{15}$N appearing in natural abundance (0.74\%).\cite{ubachs_etal1989,ubachs1997,sprengers_etal2003,vieitez_etal2007} 
This is despite the latter being of particular importance to studies of isotopic fractionation in planetary atmospheres.\cite{liang_etal2007}
Information on $f$ values and predissociation rates of the secondary isotopomers is very limited but comparison with common bands of ${}^{14}$N$_2$ shows large differences in the observed energy-level perturbations and predissociation rates.\cite{lewis_etal2005a,vieitez_etal2008a}

On the theoretical side, the seemingly erratic ordering of bands appearing in the N$_2$ XUV spectrum was finally explained by assignment to a minimal series of ${}^1\Sigma_u^+$ and ${}^1\Pi_u$ states which are mutually perturbed by strong homogeneous Rydberg-Valence interactions.\cite{carroll_collins1969,dressler1969,lefebvre-brion1969}
This basic theoretical picture culminated in a much later paper\cite{stahel_etal1983} which provided a seminal quantitative model of the N$_2$ spectrum by solution of the coupled-channel Schr\"odinger equation (CSE).\cite{mies1980a,torop_etal1987}
For this, the ${}^1\Sigma_u^+$ and ${}^1\Pi_u$ states and their homogeneous electronic-interactions were treated separately but later models have included heterogeneous coupling terms which mix the symmetry classes.\cite{edwards_etal1995,helm_etal1993}
The \emph{ab initio} calculation of electronic-transition moments connecting the excited ${}^1\Sigma_u^+$ and ${}^1\Pi_u$ states with the ground state allowed for the reproduction of the ${}^{14}$N$_2$ optical-absorption spectrum as it was then known.\cite{spelsberg_meyer2001}

A more recent CSE model built upon the previous theoretical work and more recent experimental data provided a new quantitative understanding of the energy-level structure and predissociation mechanism of ${}^1\Pi_u$ levels,\cite{lewis_etal2005a,lewis_etal2005b} including rotational and isotopic effects.
This work required the inclusion of a series of homogeneously interacting ${}^3\Pi_u$ states, some of which are dissociative, and are spin-orbit coupled to the singlet manifold.
These ${}^3\Pi_u$ levels are not accessible by optical transitions from the $X\,{}^1\Sigma_g^+$ ground state, so have been observed only sparsely and often by indirect means.\cite{sprengers_etal2005b,khakoo_etal2008,lewis_etal2008a}
An experimentally refined set of electronic-transition moments was incorporated into the CSE model\cite{haverd_etal2005} and allowed for the quantitative reproduction of experimental absolute $f$-values for many ${}^{14}$N$_2$ transitions to ${}^1\Pi_u$ excited-levels.

Fourier-transform spectroscopy (FTS) is a powerful tool which is widely used for studies at infrared and visible wavelengths, and has been extended to the vacuum-ultraviolet (VUV) by one previous instrument.\cite{thorne1991}
The principal advantages of interferometry are the combination of high resolution and the simultaneous measurement of a large spectral region, and an automatically linear frequency scale.
Recently, the technique has been extended beyond the VUV optical-transmission limit and coupled to the Desirs-beamline undulator source at the Soleil synchrotron.\cite{oliveira_etal2009,oliveira_etal2011}
This new instrument provides an unprecedented opportunity to record comprehensive ${}^{14}$N$^{15}$N absorption spectra suitable for the determination of $f$-values and natural linewidths.

\section{Experimental procedure and analysis}
\label{sec:Experimental procedure and analysis}

The characteristics of the Desirs beamline FTS\cite{oliveira_etal2009,ivanov_etal2010,oliveira_etal2011} and general properties of ultraviolet interferometry\cite{thorne1991} are given in detail elsewhere and are only briefly described here.
The spectrometer is based upon a wave-front-division interferometer and may be operated at XUV and VUV wavelengths.
A reflection-only configuration has been chosen due to the difficulty in manufacturing beam-splitters with the required optical-quality in the VUV and transparency in the XUV.
The coherent synchrotron-beam is spatially divided at the interface of two roof-shaped reflectors and an interference pattern is generated following its recombination at a photodiode.
An interferogram is recorded by systematically translating one of the reflectors, altering the optical path-difference of the divided beams.
The purified ${}^{14}\rm{N}{}^{15}\rm{N}$ sample flows continuously through a 100\,mm windowless absorption-cell with two 150\,mm\,$\times$\,28\,mm$^2$ external capillaries.
Thus, the sample gas column density could not be diagnosed absolutely.

\begin{figure*}[t]
  \centering
  \includegraphics{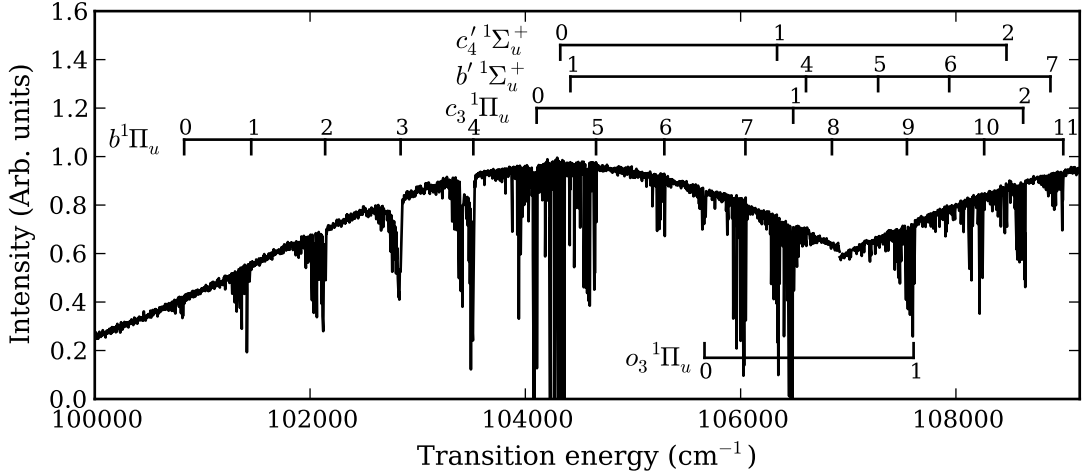}
  \caption{The measured transmission spectrum of ${}^{14}\rm{N}{}^{15}\rm{N}${}. The observed bands have been labelled according to their excited vibrational level. The plotted spectrum shows two separate measurements which are connected at 106\,900\,cm$^{-1}$.
}
  \label{fig:overall spectrum}
\end{figure*}
The synchrotron-beam spectrum was produced by an undulator insertion which provided a continuum background of high brightness and with a bandwidth of approximately 7\% of its central energy.
The energy range studied here required two overlapping spectral windows with approximately $7000$\,cm$^{-1}$ bandwidth.
Figure~\ref{fig:overall spectrum} shows a combined spectrum.
Ultimately, the FTS can achieve a resolving power close to one million for XUV wavelengths down to 40\,nm.\cite{oliveira_etal2009}
For the present study, the FTS settings have been chosen as a compromise between a good signal-to-noise ratio, resolution, and a reasonable stability of the mirror translation mechanism (see below for a detailed discussion on this latter point). 
Instrumental settings corresponding to a theoretical resolution of 0.27\,cm$^{-1}$ full-width at half-maximum (FWHM) were adopted for the presently-reported measurements.

\begin{figure*}[t]
  \centering
  \includegraphics{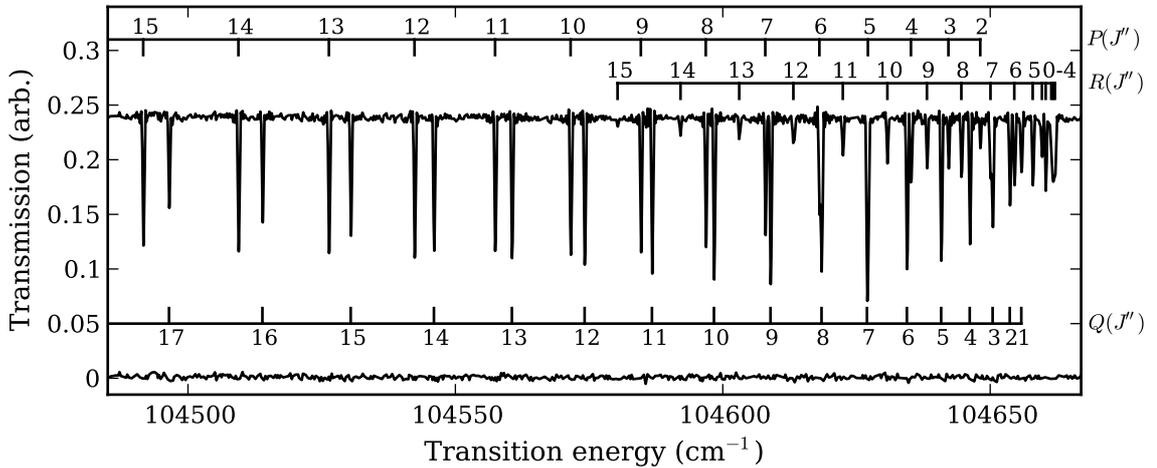}
  \caption{The recorded transmission spectrum of $b\,{}^1\Pi_u(v=5)\leftarrow X(0)$ \emph{(upper trace)} and the residual error of a model spectrum following the fitting of lineshapes \emph{(lower trace)}.}
  \label{fig:1415b05 example spectrum}
\end{figure*}
A small section of the measured spectrum is shown in Fig.~\ref{fig:1415b05 example spectrum} and is compared to a model spectrum.
The latter was constructed using fitted values for the transition energies; integrated line-strengths, $S$; and natural widths, $\Gamma$, of all observed absorption lines.
These parameters were used to synthesise an absorption cross-section, $\sigma(\nu)$, as a function of wavenumber, $\nu$, and with each line represented by a Voigt profile equivalent to a convolution of its natural Lorentzian lineshape and Gaussian-shaped Doppler broadening. 
The latter broadened the observed lines by 0.23--0.26\,cm$^{-1}$\,FWHM over the span of observed energies recorded during these room temperature measurements.
A model transmission spectrum, $I(\nu)$, was then calculated according to, 
\begin{multline}
  \label{eq:model transmission spectrum}
  I(\nu) = \biggl\lbrace I_0(\nu)\exp\bigl[ -N\sigma(\nu)\bigr]  \biggr\rbrace\scalebox{1.5}{$\ast$}\\\biggl\lbrace\text{sinc}\left(\frac{1.2\nu}{\Gamma_1}\right) \scalebox{1.5}{$\ast$} \sqrt{\frac{4\ln2}{\pi\Gamma_2^2}}\exp\left(\frac{-4\ln2\,\nu^2}{\Gamma_2^2}\right)\biggr\rbrace
\end{multline}
and directly compared to the observed transmission spectrum.
Here, $\ast$ indicates a convolution; $I_0(\nu)$ is the $\nu$-dependent intensity of the synchrotron light-source; $N$ is the column density of the target gas; and the instrumental broadening is represented by the right-hand braced term which consists of two components convolved together.
The first of these arises from the finite path-difference achievable while recording interferograms and contributes a FWHM of $\Gamma_1=0.27$\,cm$^{-1}$, which is the instrument's theoretical resolution.
The second factor contributes additional Gaussian-shaped broadening which was found to be necessary during the analysis of the ${}^{14}$N$^{15}$N spectrum.
The natural linewidths of several observed bands are known from previous experiments; i.e., excitations to  $b(0)$, $b(1)$, $b(5)$, $b(6)$, and $c_3(0)$ (with widths and references listed in Table~\ref{tab:summary 1Pi}); and $c'_4(2)$ (listed in Table~\ref{tab:summary 1Sigma}).
Without additional instrumental broadening the fitted widths of transitions to these levels were consistently larger than expected by $\sim$0.05\,cm$^{-1}$\,FWHM.
Because of this, a compensating Gaussian broadening was introduced by setting $\Gamma_2$ in Eq.~(\ref{eq:model transmission spectrum}) to 0.13\,cm$^{-1}$.
This value was arrived at by fixing the widths of model transitions to the 6 levels listed above and optimising $\Gamma_2$ so as to best match the observed spectrum.
The 6 bands were fitted independently with $\Gamma_2$ falling in the range 0.11--0.14\,cm$^{-1}$\,FWHM, neglecting a somewhat larger value for the relatively weakly appearing $b(0)-X$ band.
This spread implies an uncertainty in the adopted mean value of $\sim0.02$\,cm$^{-1}$\,FWHM.
The likely cause of this extra broadening is the non-ideal collimation of the synchrotron beam entering the FTS.
The broadening of narrow features in FTS spectra has been previously noted\cite{dooley_etal1998,learner_thorne1988} where a component of the incident radiation is not aligned with the interferometer's principal axis.
The angular distribution of the present synchrotron beam is apparently significant but is not well known.
Here, a simple Gaussian model of the resultant broadening was adopted in the absence of a more precise characterisation. 

The parameters defining each model line were adjusted by an optimisation routine until a least-squares best-fit of $I(\nu)$ of Eq.~(\ref{eq:model transmission spectrum}) to the experimental spectrum was achieved.
The deduced transition-energies were converted into excited-state term energies using the known ground-state molecular parameters of ${}^{14}\rm{N}{}^{15}\rm{N}$.\cite{bendtsen2001}
Indeed, the fitting of congested spectra was facilitated by the fixing of combination differences arising from ground-state energies for modelled $P(J'+1)$ and $R(J'-1)$ lines which terminate on common excited-state rotational levels, $J'$.
Similarly, any natural broadening of such pairs of transitions can only be due to predissociation of their common excited level, and so a common value was assumed.

The deduced integrated line strength for a transition from the $J''$ rotational level of the ground state to an excited electronic-vibrational level is equivalent to
\begin{equation}
  \label{eq:integrated line strengths}
  S_{J'J''} = \int\sigma(\nu)d\nu,
\end{equation}
where the range of integration spans the non-negligible extent of each line. 
Each measured line strength was converted into a \emph{line} $f$-value, following division by a ground-state population factor, $\alpha_{J''}$, which models a Boltzmann distribution at a temperature of 300\,K.
These have been reduced further to a set of \emph{band} $f$-values, $f_{J'}$, calculated for each rotational line by the further division of a H\"onl-London rotational line-strength factor appropriate to each rotational transition,\cite{lefebvre-brion_field2004} and an additional factor due to the degeneracy of each ground-state rotational level and the double degeneracy of ${}^1\Pi_u$ excited states. 
\begin{table}
  \centering
    \scriptsize
  \begin{tabular}{c@{\ \ \ }cc}
    \hline\hline
    Transition & Rotational & \\[-1ex]
     & branch & $\underline{S}_{J''}$ \\
    \hline
    ${}^1\Sigma\leftarrow{}^1\Sigma$ & $P$ & $J''/(2J''+1)$ \\
    ${}^1\Sigma\leftarrow{}^1\Sigma$ & $R$ & $(J''+1)/(2J''+1)$ \\
    ${}^1\Pi\leftarrow{}^1\Sigma$ & $P$ & $(J''-1)/2(2J''+1)$ \\
    ${}^1\Pi\leftarrow{}^1\Sigma$ & $Q$ & $(2J''+1)/2(2J''+1)$ \\
    ${}^1\Pi\leftarrow{}^1\Sigma$ & $R$ & $(J''+2 )/2(2J''+1)$ \\
    \hline\hline
  \end{tabular}
  \caption{Combined H\"onl-London and degeneracy factors relevant to the absorption transitions observed here.}
  \label{tab:HL and degen factors}
\end{table}
The relevant values of these combined factors, $\underline{S}_{J''}$, are listed in Table~\ref{tab:HL and degen factors}.
Then, band $f$-values were calculated according to
\begin{align}
  \label{eq:band fvalues}
  f_{J'} &= \left(\frac{4\epsilon_0 m_e c^2}{e^2}\right)\frac{S_{J'J''}}{\alpha_{J''}\underline{S}_{J''}} \notag\\[2ex]
        &= \frac{1.1296\times 10^{-4}S_{J'J''}}{\alpha_{J''}\underline{S}_{J''}},
\end{align}
where the second form is appropriate for integrated line-strengths in units of cm.
The quantity $f_{J'}$ is only representative of the integrated strength of an electronic-vibrational band if it is isolated from other electronic states.
That is, a variation of $f_{J'}$ within a rotational series indicates the presence of a perturbation,\cite{lefebvre-brion_field2004} and it is appropriate to index this by $J'$ only because the N$_2$ ground state is known to be unperturbed.
Furthermore, heterogeneous perturbations may cause $f_{J'}$ to diverge for transitions with common $J'$ but appearing within different rotational branches.\cite{vieitez_etal2008a,heays_etal2009}

The uncertainties of fitted transition-energies were estimated during the optimisation of the model spectrum and (for well-resolved lines) typically had a standard deviation of $\sim$0.02\,\,cm$^{-1}$.
An inherent property of the FTS-deduced transition energies is that they may be calibrated absolutely across an entire spectrum by the multiplication of a single scaling factor.
In principle, this could be achieved by comparison with a single reference line. 
In this case, absolute calibration was made by comparison with a previous high-precision laser-based experiment,\cite{sprengers_etal2003} with reference to 24 transitions terminating on various rotational levels of $b\,^1\Pi_u(v=5,6)$, $c_3\,^1\Pi_u(v=0)$, and $b'\,^1\Sigma^+_u(v=1)$.
The absolute uncertainty of the reference data was estimated to be 0.003\,cm$^{-1}$ and the combined statistical and systematic-calibration uncertainties of observed transition energies and deduced term values are dominated by the statistical component in most cases.
None of the laser-spectroscopic reference lines fall in the range of the higher-energy spectrum plotted in Fig.~\ref{fig:overall spectrum} and this was calibrated by reference to lines overlapping the lower-energy spectrum.

The deduced $f$-values acquired uncertainty from three different sources.
First, the random errors introduced by fitting to a noisy spectrum were assessed by the optimisation routine and varied broadly depending on the signal-to-noise ratio at each line's peak and its blendedness.
Second, the column density of flowing ${}^{14}\rm{N}{}^{15}\rm{N}$ in the FTS absorption cell is unknown so that the deduced absorption $f$-values must be uniformly scaled relative to an absolute reference.
There are no existing measurements of ${}^{14}\rm{N}{}^{15}\rm{N}$ absolute $f$-values but ${}^{14}\rm{N}_2$ has been characterised with 10\% uncertainty in a previous synchrotron- and diffraction-grating-based absorption experiment.\cite{stark_etal2005,stark_etal2008,heays_etal2009}
A calibration of ${}^{14}\rm{N}{}^{15}\rm{N}$ column density may be made relative to ${}^{14}\rm{N}_2$ assuming an isotopomeric independence of $f$-values.
This independence is reasonable only for the lowest $b\,{}^1\Pi_u$ vibrational levels because of the occurrence of large isotopomer-sensitive perturbations for higher-energy electronic-vibrational states, as discussed below.
A column density of $1.2\times10^{15}\,\text{cm}^{-2}$ was deduced for the lower-energy FTS spectrum plotted in Fig.~\ref{fig:overall spectrum} following the comparison of ${}^{14}\rm{N}_2$ and ${}^{14}\rm{N}{}^{15}\rm{N}$ $f$-values for 82 rotational transitions to 5 excited vibrational levels, $b\,^1\Pi_u(v=0-4)$.
Overlapping bands of the higher-lying spectrum were cross calibrated with this and a column density of  $1.1\times 10^{15}\,\text{cm}^{-2}$ deduced. 
Given the large number of transitions compared between experiments and spectra, the uncertainty of the deduced column densities are dominated by the original 10\% transferred from measurements of ${}^{14}$N$_2$. 
Thus, a 10\% systematic uncertainty has been assigned to all measured $f$-values.
Third, a significant addition to the uncertainty of measured $f$-values arises from environmentally-induced mechanical vibrations afflicting the translation mechanism of the FTS.
Assuming the occurrence of a periodic translation-error as the path difference of the interferometer is scanned, the resultant spectrum will include ghost reproductions of itself doubly repeated at higher and lower energies.\cite{learner_etal1996}
In the present measurements, the environmental noise is sufficiently aperiodic that sharp spectral features are not apparently aliased in this way, apart from a few extremely-weak ghost features which are likely to be associated with the very strong band $c'_4(0)\,{}^1\Sigma_u^+-X(0)$.
However, an apparent random-variation of the center energy of the synchrotron-beam bandpass resulted in completely saturated absorption lines in the recorded spectrum with minima that do not correspond to zero transmission.
Locally shifting the spectrum vertically until the lineshapes of saturated features are adequately reproduced by a model fit allowed for an assessment of this spurious effect.
In all cases, a shift of 5\% of the undulator peak-intensity, or less, was necessary and it was assumed that a worst-case vertical shift of this magnitude may affect all lines.
An additional uncertainty was estimated for each line according to the range of fitted line strength following from an assumed upward or downward shift of this magnitude.

The statistical line-fitting error and column density uncertainty are dominant for lines appearing weakly in the spectrum.
Strongly absorbed lines are more sensitive to a vertical shift of the spectrum, as are those lines near the low-intensity wings of the undulator bandpass, and this effect dominates their uncertainty.

The statistical errors of fitted natural linewidths were calculated by the line-fitting routine and an additional systematic uncertainty arises from the vertical shifting of the FTS spectrum.
The magnitude of the latter was estimated in the same fashion as the $f$-values uncertainty discussed above and, in the case of linewidths, is typically much smaller than the statistical fitting errors.
Any error in the width of instrumental broadening in Eq.~(\ref{eq:model transmission spectrum}) will lead directly to an error in fitted natural linewidths and in consideration of this an extra systematic uncertainty of $0.03$\,cm$^{-1}$\,FWHM was attributed to all fitted linewidths.
This is more conservative than the uncertainty of $\Gamma_2$ estimated above in order to account for potential model error incurred by assuming Gaussian broadening in Eq.~(\ref{eq:model transmission spectrum}).
Deduced natural linewidths nominally below $0.03$\,cm$^{-1}$\,FWHM were assumed indistinguishable from zero and neglected.

\section{CSE modelling}

The absorption $f$-values and predissociation broadening of ${}^{14}$N$^{15}$N ${}^1\Pi_u-X$ transitions were also studied here by means of a theoretical model of the molecule.
\begin{figure}[t]
  \centering
  \includegraphics{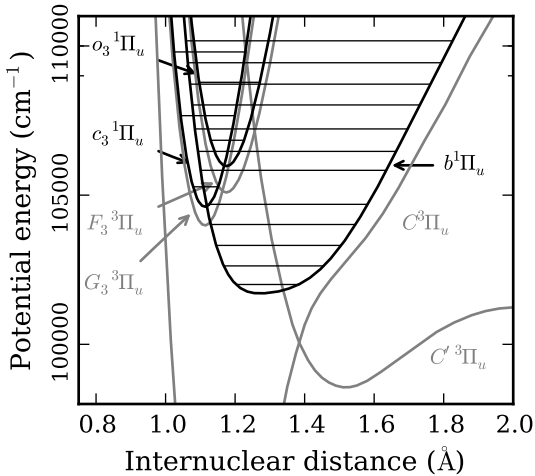}
  \caption{Model potential-energy curves for the $f$-parity ${}^1\Pi_u$ and ${}^3\Pi_u$ states of N$_2$. Horizontal lines indicate the vibrational energies of observed ${}^{14}$N$^{15}$N ${}^1\Pi_u$ levels and the energy scale is relative to the ground-state potential minimum.}
  \label{fig:fparity_potentials}
\end{figure}
A previously formulated CSE\cite{mies1980a,torop_etal1987} model of N$_2$ ${}^1\Pi_u$-states was extended to make new calculations for higher-energy ${}^{14}$N$^{15}$N vibrational levels than previously studied.\cite{lewis_etal2005a,lewis_etal2005b,haverd_etal2005}
The relevant ${}^1\Pi_u$ potential-energy curves (PECs) are plotted in Fig.~\ref{fig:fparity_potentials} and include one valence state, labelled $b$, and two Rydberg states, $c_3$ and $o_3$.
The existence of large electronic perturbations mixing these states is well known, with magnitudes of $\sim10^3\,\text{cm}^{-1}$.\cite{stahel_etal1983,spelsberg_meyer2001,lewis_etal2005a}

The Born-Oppenheimer approximation\cite{lefebvre-brion_field2004} factorises the diatomic molecular-wavefunction, $\phi_i({\bf r},R)$,  according to
\begin{equation}
  \psi_i({\bf r},R) = \chi(R)\,\phi({\bf r};R).
\end{equation}
This includes a vibrational component, $\chi(R)$, which depends only on internuclear distance, $R$, and an electronic component, $\phi({\bf r};R)$, which principally depends on  electron coordinates, ${\bf r}$, which are relative to the molecular center-of-mass.
The Born-Oppenheimer approximation is unsuitable for the modelling of high-precision experimental data because it is assumed that the electronic wavefunction has only negligible $R$-dependence.
The CSE technique considers a more general electronically-mixed wavefunction which includes contributions from all ${}^1\Pi_u$ states,
\begin{equation}
  \psi_i({\bf r},R) = \sum_{j=b,c_3,o_3}\chi_{ij}(R)\,\phi_j({\bf r};R),
  \label{eq:coupled wavefunction}
\end{equation}
where multiple independent solutions are enumerated by $i$.

The electronic wavefunctions in Eq.~(\ref{eq:coupled wavefunction}) are not explicitly calculated, instead these are parameterised by a set of diabatic PECs, $V_j(R)$, and $R$-dependent electronic-coupling parameters, $\left<i|H^\text{el}(R)|j\right>$, which may be arranged into a convenient matrix,
\begin{equation}
{\bf V}(R) = \begin{pmatrix}
V_b(R)                           & \bigl<b\bigl|H^\text{el}(R)\bigr|c_3\bigr> & \bigl<b\bigl|H^\text{el}(R)\bigr|o_3\bigr> \\
\bigl<b\bigl|H^\text{el}(R)\bigr|c_3\bigr> & V_{c_3}(R)                         & \bigl<c_3\bigl|H^\text{el}(R)\bigr|o_3\bigr> \\
\bigl<b\bigl|H^\text{el}(R)\bigr|o_3\bigr> & \bigl<c_3\bigl|H^\text{el}(R)\bigr|o_3\bigr> & V_{o_3}(R)
\label{eq:Vmatrix}
\end{pmatrix}.
\end{equation}
Then, a matrix of coupled vibrational-wavefunctions is the solution of the radial coupled-channel Schr\"odinger equation,
\begin{equation}
\frac{d^2}{dR^2}{\boldsymbol\chi}(R) = \frac{-2\mu}{\hbar^2}{\boldsymbol\chi}(R)\left[E-{\bf V}(R)\right].
\label{eq:coupled radial equation}
\end{equation}
Separate solutions may be calculated for each value of the total energy, $E$, and reduced mass, $\mu$.
Adjustment of $\mu$ is sufficient to extend the CSE calculations to any isotopomer of N$_2$ without alteration of the mass-independent PECs and coupling parameters.

The ${}^1\Pi_u$ PECs and interaction parameters were optimised with comparison to a database of experimental energy levels.\cite{lewis_etal2005a}
The addition of an extra term, ${\hbar^2J(J+1)}/{2\mu R^2}$, representing a centrifugal potential was added to the CSE PECs\cite{lewis_etal2005b,haverd_etal2005} which allowed for the calculation of excited rotational levels, according to the total angular-momentum quantum-number, $J$. 
Previously, the CSE model was used to calculate term origins and rotational constants for all ${}^1\Pi_u$ levels below 105\,350\,cm$^{-1}$ and reproduced their experimentally known values to within 0.5 and 0.007\,cm$^{-1}$, respectively, for all three isotopomers of N$_2$.\cite{lewis_etal2005a}
Similar agreement is seen for calculations compared to the newly observed ${}^{14}$N$^{15}$N levels following a slight adjustment of the ${}^1\Pi_u$ PECs necessary to account for the present extension to higher energy.

The additional definition of a ground-state PEC and vibrational wavefunction $\chi_X(R)$, and $R$-dependent electronic-transition moments,\cite{haverd_etal2005} $R^e_{jX}(R)$, for the optical transitions $b-X$, $c_3-X$, and $o_3-X$ permitted the calculation of an absorption cross-section, $\sigma_{iXJ'J''}$ from the CSE-calculated coupled wavefunctions, according to
\begin{multline}
  \sigma_{iXJ'J''} = \\ \frac{\pi\nu}{4\hbar\epsilon_0}\left| \sum_{j=b,c_3,o_3} S_{iXJ'J''}\int\chi_{ij}^\dagger(R)\ R^e_{jX}(R)\ \chi_{\tiny X}(R)\,dR\,\right|^2.
  \label{eq:CSE cross section}
\end{multline}
Here, $S_{iXJ''J'}$ is a H\"onl-London rotational line-strength factor\cite{herzberg1989} which differs for electronic states of different symmetry and ground- and excited-state rotational levels, $J'$ and $J''$, respectively.
Resonances in the $\sigma_{iXJ'J''}$ spectrum are of mixed electronic character and must be assigned to approximate electronic-vibrational progressions as is done for experimental spectra.
The calculated cross-sections may be converted to equivalent band $f$-values and the model electronic-transition moments were optimised with respect to a set of experimental ${}^{14}$N$_2$ absorption $f$-values.\cite{haverd_etal2005}
These transition moments were not altered for the present calculations of ${}^{14}$N$^{15}$N.

Predissociation of the ${}^1\Pi_u$ states was long suspected to arise from multiple spin-orbit perturbations with bound and dissociative states of ${}^3\Pi_u$ symmetry.\cite{carroll_collins1969,dressler1969,ubachs_etal1989}
The precise mechanism has been deduced by Lewis \emph{et al.}\cite{lewis_etal2005a,lewis_etal2008a,lewis_etal2008b} and the resultant picture of ${}^3\Pi_u$ states consists of two with valence character, $C'\,{}^3\Pi_u$ and $C\,{}^3\Pi_u$,  and two Rydberg-states, $G_3\,{}^3\Pi_u$ and $F_3\,{}^3\Pi_u$; with PECs plotted in Fig.~\ref{fig:fparity_potentials}.
The $C'\,{}^3\Pi_u$ and $C\,{}^3\Pi_u$ states were incorporated into an extended matrix of the form of Eq.~(\ref{eq:Vmatrix}), along with an electronic coupling between them and spin-orbit interactions mixing them with the ${}^1\Pi_u$ states.\cite{lewis_etal2005a,lewis_etal2005b,haverd_etal2005}
This allowed for quantitative CSE modelling of the predissociation linewidths of the lowest ${}^1\Pi_u$ levels.
These calculations demonstrated good agreement with the then-available experimental linewidths collected from all three isotopomers of N$_2$.\cite{lewis_etal2005a,lewis_etal2008a}
The inclusion of an unbound state, $C'\,{}^3\Pi_u$, in the CSE formulation permits the calculation of Eq.~(\ref{eq:CSE cross section}) for all values of $E$, in which case the predissociation of ${}^1\Pi_u$ levels manifests as broadened resonances in a continuous absorption cross-section.

The CSE model employed here does not include states with ${}^1\Sigma_u^+$ symmetry, and so is only strictly representative of the $f$-parity levels of the doubly-degenerate ${}^1\Pi_u$ states.
This results from a rotational-perturbation which mixes $e$-parity levels of the ${}^1\Pi_u$ and ${}^1\Sigma_u^+$ states,\cite{lefebvre-brion_field2004} and increases in severity approximately proportionately to $J(J+1)$.
Consequently, the lowest-$J$ $e$-parity ${}^1\Pi_u$ rotational-levels will be least perturbed, as will those for ${}^1\Pi_u$ vibrational-levels which lie below the onset of the ${}^1\Sigma_u^+$ states.
Because of these rotational interactions, the ${}^{14}$N$^{15}$N linewidths and $f$-values deduced from observed $P$- or $R$-branch transitions with $e$-parity ${}^1\Pi_u$ excited-states may then differ from those inferred from $Q$-branch lines with excited levels of $f$-parity.
All of the strictly-$e$-parity ${}^1\Sigma_u^+$ levels will be more or less heterogeneously perturbed; except those with $J=0$, because of the lack of perturbing ${}^1\Pi_u$ $J=0$ levels.

\section{Results and discussion}

A total of 25 absorption bands were observed and reduced to a set of transition energies, line strengths, and natural linewidths.
Excited-state term values have been calculated from the observed transition energies and known ground-state level energies,\cite{bendtsen2001} and band $f$-values have been calculated from the observed line strengths according to Eq.~(\ref{eq:band fvalues}).
The complete set of deduced line parameters is available in an on-line data archive accompanying this article.\cite{heays_etal2011_data_archive} 
Important features of the observed ${}^1\Pi_u-X$- and ${}^1\Sigma_u^+-X$- bands are discussed in the following subsections.
This discussion does not emphasise classical spectroscopic analysis of band parameters and molecular constants, and instead favours a direct assessment of the primary data.
This is believed to be appropriate because the highly perturbed N$_2$ spectrum can be more profitably modelled by CSE-type calculations than local deperturbation methods.

Evident in the observations of excited ${}^1\Pi_u$ and ${}^1\Sigma_u^+$ states are a number of rotationally-dependent transition-energy and linewidth perturbations that indicate the presence of additional ${}^3\Pi_u$ states beyond the scope of the present CSE model. 
This includes the appearance of extra lines in the spectrum arising from nominally forbidden absorption to these levels.
The analysis and discussion of these perturbations is reserved for a later publication. 

\subsection{$\boldsymbol{{}^1\Pi_u}$ states}
\label{sec:results 1Pi states}

\begin{table*}[t]
\begin{minipage}{\textwidth}
  \centering
    \scriptsize
  \begin{tabular}{ccclllllll}

\hline\hline
Excited state&\multicolumn{1}{c}{  $N$\footnote[1]{The number of observed rotational transitions.} }&\multicolumn{1}{c}{  $J'_\text{max}$\footnote[2]{The maximum observed rotational excitation.} }&\multicolumn{1}{c}{  $T_0$      }&\multicolumn{1}{c}{ $T_0^\text{ref}$ }&\multicolumn{1}{c}{  $f_0$        }&\multicolumn{1}{c}{  $f_0^\text{CSE}$  }&\multicolumn{1}{c}{  $\Gamma_0$  }&\multicolumn{1}{c}{$\Gamma_0^\text{CSE}$}& \multicolumn{1}{c}{  $\Gamma_0^\text{ref}$}\\
\noalign{\vspace{1ex}}\hline

$b(0)$   & 39 & 14 & 100\,830.42                                          & 100\,830.42\footnote[3]{Ref. \onlinecite{sprengers_etal2005}.} & 0.0023(3)             & 0.0023  & 0.06(3) & 0.068  & 0.065(21)\footnotemark[3]   \\
$b(1)$   & 61 & 22 & 101\,453.68                                          & 101\,453.24\footnotemark[3]                                         & 0.0086(9)             & 0.0084  & 0.04(3) & 0.030  & 0.028(16)\footnotemark[3]   \\
$b(2)$   & 63 & 22 & 102\,140.36                                          &                                                                     & 0.021(2)              & 0.020   & 0.76(3) & 0.69   &                             \\
$b(3)$   & 61 & 22 & 102\,841.46                                          &                                                                     & 0.041(6)              & 0.039   & 2.66(3) & 2.60   &                             \\
$b(4)$   & 71 & 25 & 103\,517.22                                          &                                                                     & 0.063(7)              & 0.058   & 1.46(4) & 1.23   &                             \\
$c_3(0)$ & 80 & 28 & 104\,104.93                                          & 104\,104.93\footnote[4]{Ref. \onlinecite{sprengers_etal2003}.} & 0.045(5)              & 0.047   & 0.12(3) & 0.094  & 0.11(3)\footnotemark[3]     \\
$b(5)$   & 63 & 24 & 104\,656.94                                          & 104\,656.90\footnotemark[4]                                         & 0.0056(6)             & 0.0049  &         & 0.010  & $<$0.006\footnotemark[3]    \\
$b(6)$   & 50 & 19 & 105\,291.22                                          & 105\,291.23\footnotemark[4]                                         & 0.0035(4)             & 0.0033  &         & 0.0085 & 0.0074(29)\footnotemark[3]  \\
$o_3(0)$ & 50 & 20 & 105\,664.43                                          &                                                                     & 0.00059(7)            & 0.00064 &         &        &                             \\
$b(7)$   & 65 & 23 & 106\,044.83                                          & 106\,045.02\footnotemark[4]                                         & 0.016(2)              & 0.016   &         &        &                             \\
$c_3(1)$ & 67 & 25 & 106\,488.53                                          &                                                                     & 0.039(5)              & 0.040   & 0.03(3) &        &                             \\
$b(8)$   & 21 & 21 & 106\,848.07\footnote[5]{Predicted by the CSE model.} &                                                                     & $\!\!\!\!\!<\!0.0003$ & 0.00013 &         &        &                             \\
$b(9)$   & 53 & 19 & 107\,545.05                                          & 107\,545.26\footnote[6]{Ref. \onlinecite{vieitez_etal2007}.}   & 0.0041(5)             & 0.0035  & 0.10(3) &        &                             \\
$o_3(1)$ & 64 & 23 & 107\,606.72                                          & 107\,607.03\footnotemark[6]                                         & 0.015(2)              & 0.014   & 0.25(3) &        &                             \\
$b(10)$  & 60 & 22 & 108\,262.65                                          &                                                                     & 0.010(1)              & 0.0092  & 0.10(3) &        &                             \\
$c_3(2)$ & 65 & 25 & 108\,624.65                                          & 100\,624.69\footnote[7]{Ref. \onlinecite{vieitez_etal2008a}.}  & 0.012(1)              & 0.011   & 0.14(3) &        &                             \\
$b(11)$  & 47 & 17 & 108\,997.53                                          &                                                                     & 0.0042(5)             & 0.0037  & 0.30(3) &        &                             \\

\noalign{\vspace{1ex}}\hline\hline

  \end{tabular}
  \caption{Summary of deduced line parameters extrapolated to $J'=0$ for excited ${}^1\Pi_u$ states. A term origin, $T_0$, band $f$-value, $f_0$, and natural linewidth, $\Gamma_0$, is given for each absorption band. Comparable quantities are also listed as determined from the CSE model and previous observations, superscripted with ``CSE'' and ``ref'', respectively.}
  \label{tab:summary 1Pi}
\end{minipage}
\end{table*}
A summary set of \emph{rotationless} quantities is given in Table~\ref{tab:summary 1Pi} for the observed ${}^1\Pi_u-X$ transitions.
For these, the experimentally deduced line parameters have been extrapolated to $J'=0$ in order to facilitate a simple comparison with CSE-calculated values and other experiments. 
There is no physical level corresponding to $J'=0$ in the case of ${}^1\Pi_u$ states but a smooth extrapolation has been made to a virtual $J'=0$ level which is, in principle, free of ${}^1\Sigma_u^+\sim{}^1\Pi_u$ rotational mixing for both $e$- and $f$-parity levels.
Also listed are CSE calculations of ${}^1\Pi_u$ rotationless $f$-values and predissociation linewidths, as well as any previous linewidth measurements.

\begin{figure*}[t]
  \centering
  \includegraphics{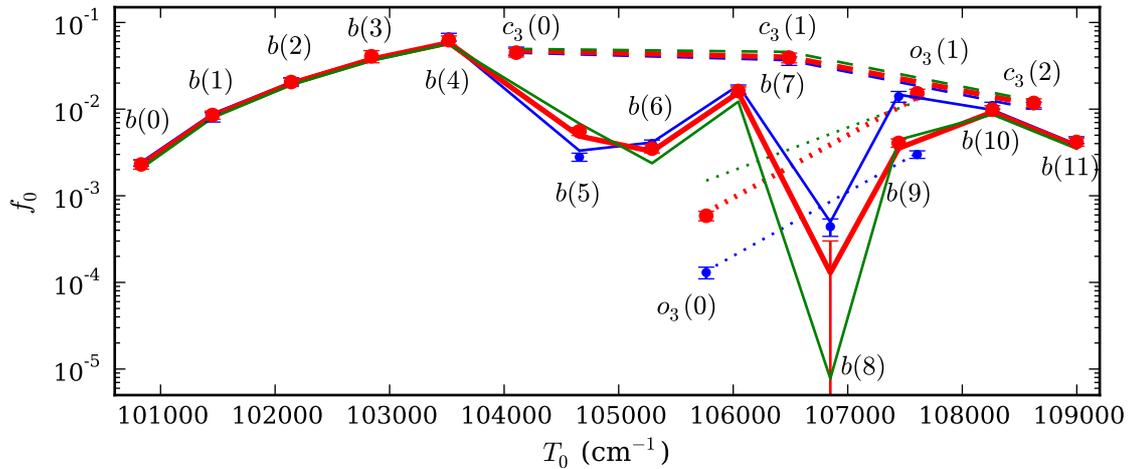}
  \caption{
Band $f$-values, $f_0$, extrapolated to $J'=0$ for the ${}^1\Pi_u\leftarrow X$ bands of ${}^{14}$N$_2$,  ${}^{14}$N$^{15}$N, and ${}^{15}$N$_2$ (\emph{blue, red, and green lines and circles; respectively}); plotted versus their ${}^{14}$N$^{15}$N term-origins, $T_0$.
Experimental data from the present ${}^{14}$N$^{15}$N observations and previous ${}^{14}$N$_2$ measurements\cite{stark_etal2005,stark_etal2008,heays_etal2009} are plotted as filled circles, and CSE-calculated $f$-values are represented by line vertices joining the vibrational series of $b\,{}^1\Pi_u$, $c_3\,{}^1\Pi_u$, and $o_3\,{}^1\Pi_u$ \emph{(solid, dashed, and dotted lines; respectively)}.
    The locations of $o_3(1)$ and $b(9)$ have been shifted from their true $T_0$ for clarity and only an upper-bound could be determined for the rotationless $f$-value of $b(8)-X$.
}
  \label{fig:f0 summary 1Piu}
\end{figure*}
Figure~\ref{fig:f0 summary 1Piu} compares the available experimental ${}^1\Pi_u-X$ $f$-values for ${}^{14}$N$_2$ and ${}^{14}$N$^{15}$N with CSE calculations for all isotopomers. 
It was assumed during the calibration of the experimental column density that the $f$-values of $b(v=0-4)-X$ are mass independent, and this is supported by the model calculations.
In general, the CSE model predictions of ${}^{14}$N$^{15}$N $f$-values are in excellent agreement with the new experimental values deduced here, validating the accuracy of the model electronic-transition moments constructed with reference to ${}^{14}$N$_2$ only.\cite{haverd_etal2005} 

The mass dependence of $b\,{}^1\Pi_u-X$ $f$-values is much greater for vibrational levels above the onset of the ${}^1\Pi_u$ Rydberg states.
Large variations with isotopomer are evident for absorption to $b(v=5,6)$ and $o_3(v=0,1)$, and the largest of all predicted for $b(8)-X$.
No rotational lines from this band with $J'<12$ were observed in the present ${}^{14}$N$^{15}$N spectrum and only an upper bound could be deduced for the $J=0$ $f$-value, in agreement with its CSE-predicted weakness.

Particularly interesting is the factor-of-5 difference in the rotationless $f$-values of $b(9)-X$ and $o_3(1)-X$, with absorption to $b(9)$ being the stronger in observations of ${}^{14}$N$_2$\cite{vieitez_etal2007} and $o_3(1)$ in the present spectrum of ${}^{14}$N$^{15}$N.
This switch occurs because of a change of sign in the quantum interference of mixed  $b-X$ and $o_3-X$ transition amplitudes where there is a change in energy ordering.
That is, the term origin of $o_3(1)$ lies below $b(9)$ for ${}^{14}$N$_2$ and above it for the other isotopomers.
Indeed, there is a level crossing in the ${}^{14}$N$_2$ rotational series of $o_3(1)$ and $b(9)$ between $J=4$ and 5, which leads to the $o_3(1)-X$ $f$-value being the larger for this isotopomer also for $J\geq5$.\cite{vieitez_etal2007}
These isotopomeric differences are well modelled by the nonlinear sum of transition amplitudes in Eq.~(\ref{eq:CSE cross section}), whereby mass-dependent shifts of ${}^1\Pi_u$ levels are sufficient to alter their degree of mixing and cause the observed variation of $f$-values to occur in the CSE calculations.

\begin{figure*}[t]
  \centering
  \includegraphics{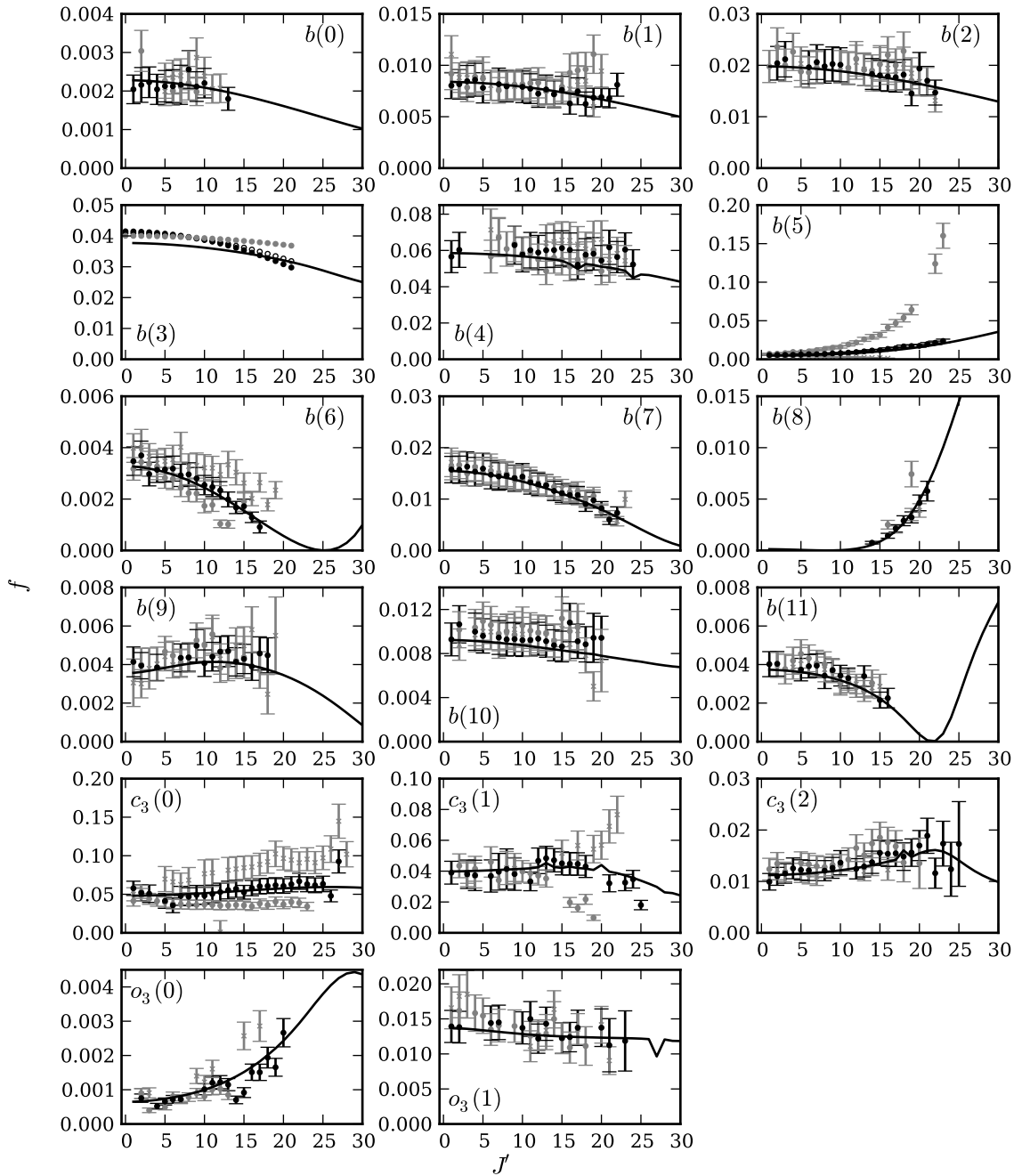}
  \caption{
    Rotational dependence of ${}^1\Pi_u-X$ absorption $f$-values, labelled according to excited level.
    Values are given for experimental and modelled $Q$-branch transitions \emph{(dark points and lines, respectively)}, as well as for experimental $P$- and $R$-branch transitions \emph{(light circles and crosses, respectively)}.
    In the case of $b(3)-X$, the experimental $f$-values were constrained to simple polynomial forms for each of the $P$, $Q$, and $R$ branches \emph{(light, empty, and filled circles; respectively)}.
The uncertainty of $b(3)-X$ $f$-values is estimated to be 0.006.
}
  \label{fig:1Piu_fJ}
\end{figure*}
The rotational dependence of ${}^1\Pi_u-X$ $f$-values is summarised in the various plots of Fig.~\ref{fig:1Piu_fJ}.
The experimentally determined and CSE-modelled $f$-values are in excellent agreement for all of the observed vibrational bands, even where the variation with $J$ approaches an order of magnitude, as is the case for absorption to $b(6)$, $b(8)$, and $o_3(0)$.
The model actually predicts the complete disappearance of $b(6)-X$ and $b(11)-X$ for rotational transitions above those observed.
Additionally, the band head of $b(8)-X$ is completely invisible in the FTS spectrum and the experimental $f$-values that have been deduced exhibit a rapid growth, in line with the CSE calculations.
In fact, the CSE model predicts completely-destructive interference of the ${}^1\Pi_u-X$ transition amplitudes describing absorption to $b(8)$, leading to a minimum in the ${}^{14}$N$^{15}$N $Q$-branch $f$-values at $J'=8$, and similar minima at $J'=12$ and 0 for the cases of ${}^{14}$N$_2$ and ${}^{15}$N$_2$.

In the case of $b(3)-X$, it was not possible to independently fit lineshapes to each rotational transition because their linewidths are similar to their spacing, and the $P$ and $Q$ branches are unfortuitously overlapped.
Instead, a smooth variation of band $f$-values was assumed for each rotational branch and constrained to the following polynomials with optimised coefficients:
\begin{align*}
f_P(J') &= 0.042 - 2.5\times 10^{-5}J'(J'+1);\\
f_Q(J') &= 0.041 - 2.0\times 10^{-5}J'(J'+1);\\
f_R(J') &= 0.040 - 0.7\times 10^{-5}J'(J'+1).\\
\end{align*}
An attempt to model the $f$-values of all three $b(3)-X$ rotational branches to a common function resulted in a poorer fit to the FTS spectrum, suggesting that the deduced splitting of the branches for high $J$ is a real phenomenon.

Several ${}^1\Pi_u-X$ bands are observed to have $P$- and $R$-branch $f$-values which increasingly diverge from the $Q$-branch with increasing $J$, notably for excitation to $b(5)$, $b(6)$, $c_3(0)$, $c_3(1)$, and $o_3(0)$.
This is due to heterogeneous mixing with the ${}^1\Sigma_u^+$ states and is discussed further in Sec.~\ref{sec:results 1Sigma states}.

\begin{figure*}[t]
  \centering
  \includegraphics{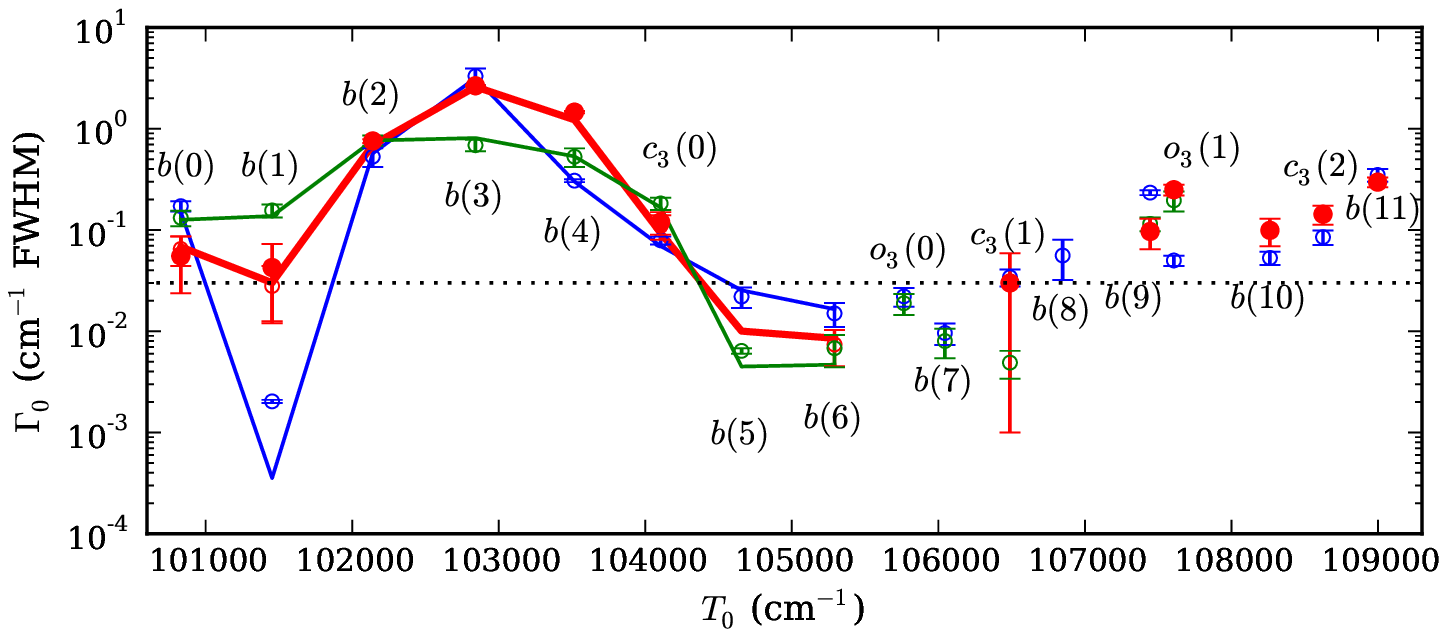}
  \caption{
    Experimental natural linewidths, $\Gamma_0$, extrapolated to $J'=0$ of excited ${}^1\Pi_u$-states of ${}^{14}$N$_2$, ${}^{14}$N$^{15}$N, and ${}^{15}$N$_2$ (\emph{blue, red, and green lines and points; respectively}); plotted versus their ${}^{14}$N$^{15}$N term-origins, $T_0$.
    Experimentally determined natural linewidths are from the present and previous observations \emph{(filled and empty circles, respectively)}, and CSE-calculated predissociation linewidths (which exclude radiative broadening) are connected by lines.
    The locations of $o_3(1)$ and $b(9)$ have been shifted from their true $T_0$ for clarity and a horizontal line has been superimposed indicating the minimum width detectable in the present ${}^{14}$N$^{15}$N spectrum.
    References to previous measurements of ${}^{14}$N$^{15}$N linewidths are given in Table~\ref{tab:summary 1Pi}, values for ${}^{14}$N$_2$ and ${}^{15}$N$_2$ are collected from a variety of sources.\cite{lewis_etal2008a,sprengers_etal2004a,sprengers_etal2004b,sprengers_etal2005,sprengers_ubachs2006,stark_etal2008,ubachs_etal1989,ubachs_etal2000,vieitez_etal2007}
    }
  \label{fig:1Piu_Gamma0}
\end{figure*}
The experimental and modelled rotationless linewidths of ${}^{14}$N$^{15}$N span several orders-of-magnitude, as demonstrated in Fig.~\ref{fig:1Piu_Gamma0}.
CSE-calculated and previous experimental linewidths of ${}^{14}$N$_2$ and ${}^{15}$N$_2$ are also shown in this figure and exhibit a large mass dependence for many ${}^1\Pi_u$ levels.
The various experimental natural linewidths are not strictly comparable with the CSE-calculated dissociation linewidths which do not include the broadening influence of radiative decay. 
In effect, this is only significant for the least predissociated levels such as $b(1)$, where the difference between calculated and observed width for ${}^{14}$N$_2$ $b(1)$ is eliminated once radiative decay is considered.\cite{lewis_etal2005a}

In Fig.~\ref{fig:1Piu_Gamma0}, the CSE model correctly reproduces the large ${}^{14}$N$^{15}$N linewidths of $b(v=2,3,4)$ and $c_3(0)$, all of which differ by a significant amount from the previously observed ${}^{14}$N$_2$ and ${}^{15}$N$_2$ widths, providing further validation of its predictiveness.
The lack of observable linewidths encountered in the ${}^{14}$N$^{15}$N spectrum for excited levels between $b(5)$ and $b(8)$ is unsurprising given the narrowness of their linewidths predicted by the CSE calculations and observed for the other isotopomers.
A broadening of ${}^{14}$N$^{15}$N levels then occurs for $b(9)$ and higher levels, which is consistent with the general trends observed for ${}^{14}$N$_2$ and ${}^{15}$N$_2$.
There is a reversal of the ratio of $b(9)$ and $o_3(1)$ linewidths with respect to ${}^{14}$N$_2$ and the other isotopomers apparent in Fig.~\ref{fig:1Piu_Gamma0} which is analogous to the previously-discussed $f$-value anomalies of $b(9)-X$ and $o_3(1)-X$.

The general trend of predissociation widths in Fig.~\ref{fig:1Piu_Gamma0} suggests a maximum near 103\,000\,cm$^{-1}$ and minimum around 106\,000\,cm$^{-1}$.
This is consistent with the picture of ${}^3\Pi_u$ predissociation deduced by a previous CSE model\cite{lewis_etal2008b} which included a representation of not only $C\,{}^3\Pi_u$ and $C'\,{}^3\Pi_u$ but also the Rydberg states $G_3\,{}^3\Pi_u$ and $F_3\,{}^3\Pi_u$, a lack of which limits the present model to the calculation of widths for levels below $o_3(0)$.\cite{lewis_etal2005a}
No ${}^1\Pi_u$ states were included in the $G_3$/$F_3$ model but it is reasonable to assume an approximate transfer of the energy dependence of ${}^3\Pi_u$ predissociation to these, explaining the broad similarity of ${}^1\Pi_u$ linewidths in Fig.~\ref{fig:1Piu_Gamma0} and the predicted widths of ${}^3\Pi_u$ states shown in Fig.~4 of Ref.~\onlinecite{lewis_etal2008b}.

\begin{figure*}[t]
  \centering
  \includegraphics{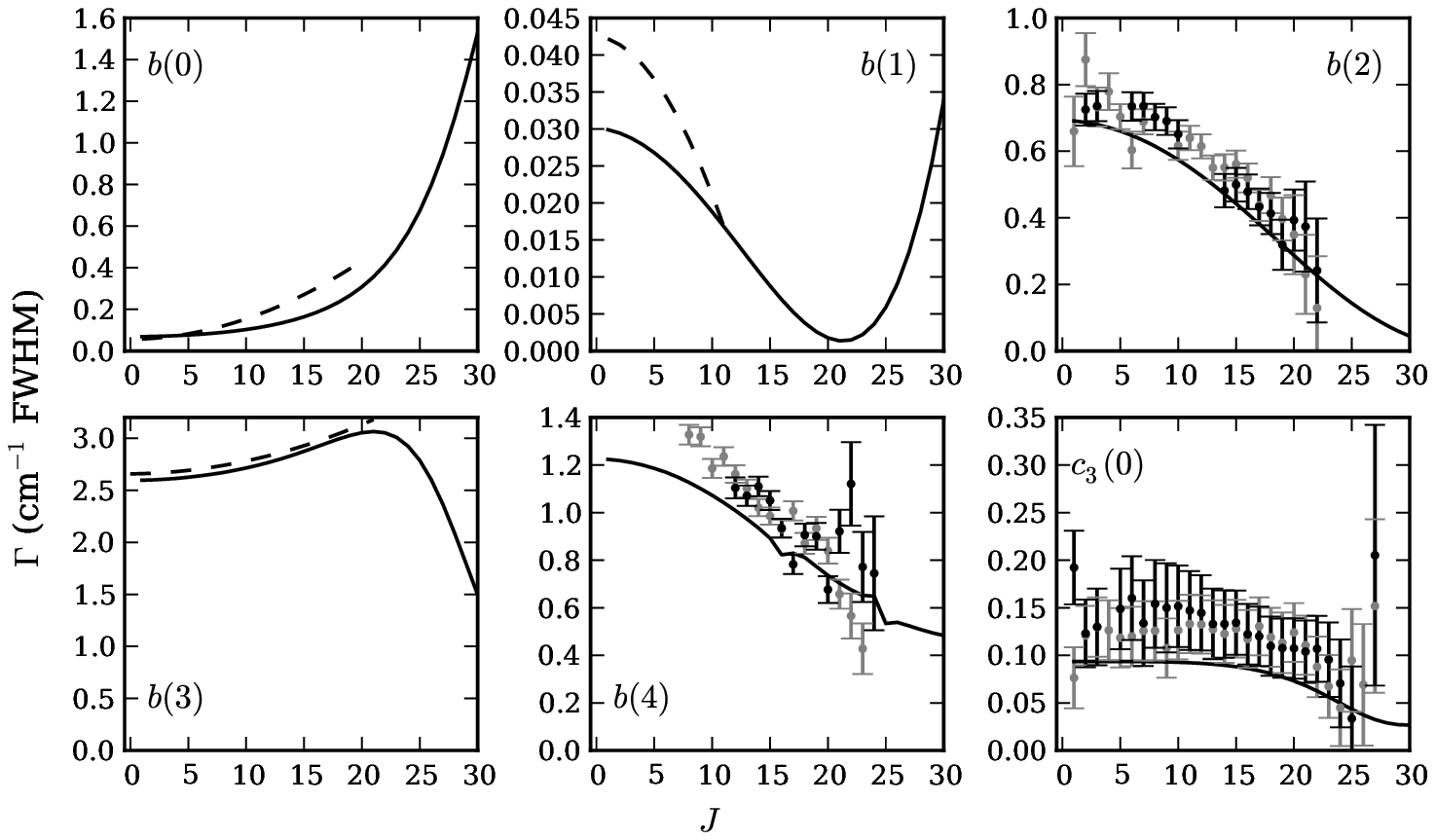}
  \caption{
    Linewidths of ${}^1\Pi_u$ levels.
    Values are given for experimental and CSE-modelled $f$-parity levels \emph{(dark circles and lines, respectively)} and experimental $e$-parity levels \emph{(light circles)}.
    Some $f$-parity experimental widths were deduced by assuming a simple polynomial form \emph{(dashed lines)} and the uncertainties of these are dominated by the estimated 0.03\,cm$^{-1}$\,FWHM systematic error due to the necessary calibration of the FTS instrument-function.
  }
  \label{fig:1Piu_GammaJ}
\end{figure*}
The rotational dependencies of comparable observed and calculated ${}^1\Pi_u$ linewidths are plotted in Fig.~\ref{fig:1Piu_GammaJ}.
The linewidths of some excited levels were assumed to follow a simple polynomial dependence with fitted coefficients.
This was necessary in the case of $b(0)-X$ because this band appears relatively weakly in the spectrum and with a poor signal-to-noise ratio, whereas the linewidths of $b(1)-X$ are close to the lower bound of measurability, and the rotational lines of $b(3)-X$ are excessively blended.
For these bands, fitted natural linewidths were assumed to have the following forms for both $e$- and $f$-parity levels:
\begin{align*}
  \Gamma_{b(0)}(J) &= 0.055 + 9.0\times10^{-4}\,J(J+1); \\
  \Gamma_{b(1)}(J) &= 0.043 - 1.9\times10^{-4}\,J(J+1); \\
  \Gamma_{b(3)}(J) &= 2.66 + 1.1\times10^{-3}\,J(J+1)\ \ \,\text{cm}^{-1}\,\text{FWHM}.\\
\end{align*}
The similarity of $e$- and $f$-parity linewidths is a reasonable assumption given the negligible heterogeneous mixing of $b(v=0,1,3)$ with the higher-lying ${}^1\Sigma_u^+$ states, evident in the $f$-values of Fig.~\ref{fig:1Piu_fJ}.

The agreement between CSE-calculated and experimental linewidths in Fig.~\ref{fig:1Piu_GammaJ} is very good even for the $J$-dependent trends of $b(2)$ and $b(4)$ widths, which exhibit variation up to a factor of 4.
There is, however, an apparently-uniform under-calculation of linewidths for $b(2)$, $b(4)$ and $c_3(0)$ which cannot be explained by the neglect of additional broadening due to radiative decay.
This suggests that an adjustment of the CSE coupling of ${}^1\Pi_u$ and ${}^3\Pi_u$ states may be necessary.

\subsection{$\boldsymbol{{}^1\Sigma^+_u}$ states}
\label{sec:results 1Sigma states}

\begin{table*}[t]
\begin{minipage}{\textwidth}
  \centering
  \scriptsize
  \begin{tabular}{ccclllll}

\hline\hline
Excited state&\multicolumn{1}{c}{  $N$\footnote[1]{The number of observed rotational transitions.} }&\multicolumn{1}{c}{  $J'_\text{max}$\footnote[2]{The maximum observed rotational excitation.} }&\multicolumn{1}{c}{  $T_0$      }&\multicolumn{1}{c}{ $T_0^\text{ref}$ }&\multicolumn{1}{c}{  $f_0$        }&\multicolumn{1}{c}{  $\Gamma_0$  }& \multicolumn{1}{c}{  $\Gamma_0^\text{ref}$}\\
\noalign{\vspace{1ex}}\hline

$c'_4(0)$ & 53 & 27 & 104\,324.46 & 104\,324.64\footnote[3]{An unpublished observation in high pressure ${}^{14}$N$_2$ absorption measurements of Ref. \onlinecite{stark_etal2005}.} & 0.134(14)   &         &                                                                \\
$b'(1)$   & 28 & 17 & 104\,419.12 & 104\,419.01\footnote[4]{Ref. \onlinecite{sprengers_etal2003}.}                                                                                   & 0.0003(2)   &         &                                                                \\
$c'_4(1)$ & 37 & 19 & 106\,338.46 &                                                                                                                                                     & 0.0060(7)   &         & 0.022(3)\footnote[5]{Ref. \onlinecite{ubachs_etal2001}.}  \\
$b'(4)$   & 34 & 22 & 106\,607.60 &                                                                                                                                                     & 0.0017(3)   &         &                                                                \\
$b'(5)$   & 34 & 17 & 107\,277.10 &                                                                                                                                                     & 0.00099(13) &         &                                                                \\
$b'(6)$   & 35 & 19 & 107\,937.08 & 107\,937.50\footnote[6]{Ref. \onlinecite{vieitez_etal2007}.}                                                                                     & 0.0015(2)   & 0.08(3) &                                                                \\
$c'_4(2)$ & 37 & 21 & 108\,469.71 &                                                                                                                                                     & 0.0011(2)   &         &                                                                \\
$b'(7)$   & 22 & 17 & 108\,877.93 & 108\,877.89\footnote[7]{Ref. \onlinecite{vieitez_etal2008a}.}                                                                                    & 0.00042(8)  & 0.11(4) &                                                                \\

\noalign{\vspace{1ex}}\hline\hline

  \end{tabular}
  \caption{Summary of deduced line parameters extrapolated to $J'=0$ for excited ${}^1\Sigma_u^+$ states. A term origin, $T_0$, band $f$-value, $f_0$, and natural linewidth, $\Gamma_0$, is listed for each absorption band. Comparable quantities determined from previous observations are superscripted with ``ref''.}
  \label{tab:summary 1Sigma}
\end{minipage}
\end{table*}
A summary set of rotationless quantities is given in Table~\ref{tab:summary 1Sigma} for the observed ${}^1\Sigma_u^+-X$ transitions.
There are no comparable experimental $f$-values or rotationless natural-linewidths apart from a measurement of the width of $c'_4(1)$\cite{ubachs_etal2001} which falls below the lower measurable limit of the present observations, as do the linewidths of all other observed ${}^1\Sigma_u^+-X$ transitions with the exception of excitations to $b'(6)$ and $b'(7)$.
No $J$-dependence of linewidths was observed for these two vibrational levels.

The dearth of ${}^1\Sigma_u^+$ states observed to predissociate is unsurprising if it is assumed that this should follow from a highly indirect mechanism requiring their rotational coupling to ${}^1\Pi_u$ states which are in turn spin-orbit coupled to ${}^3\Pi_u$ states.
Interestingly, the approximate $J(J+1)$ proportionality expected for rotational-coupling-induced predissociation is not observed for $b'(6)$ and $b'(7)$.
An alternative homogeneous-predissociation mechanism involving a direct interaction of $b'\,{}^1\Sigma_u^+$ with the $\Omega=0$ levels of $C'\,{}^3\Pi_u$ or $C\,{}^3\Pi_u$ offers a possible explanation of the $b'(6)$ and $b'(7)$ linewidths.
Significantly $J$-dependent natural linewidths were observed for $b'(4)$ for $J\geq 3$ and will be discussed in a future paper concerning the detection of ${}^3\Pi_u$ states in the ${}^{14}$N$^{15}$N spectrum.

\begin{figure*}[t]
  \centering
  \includegraphics{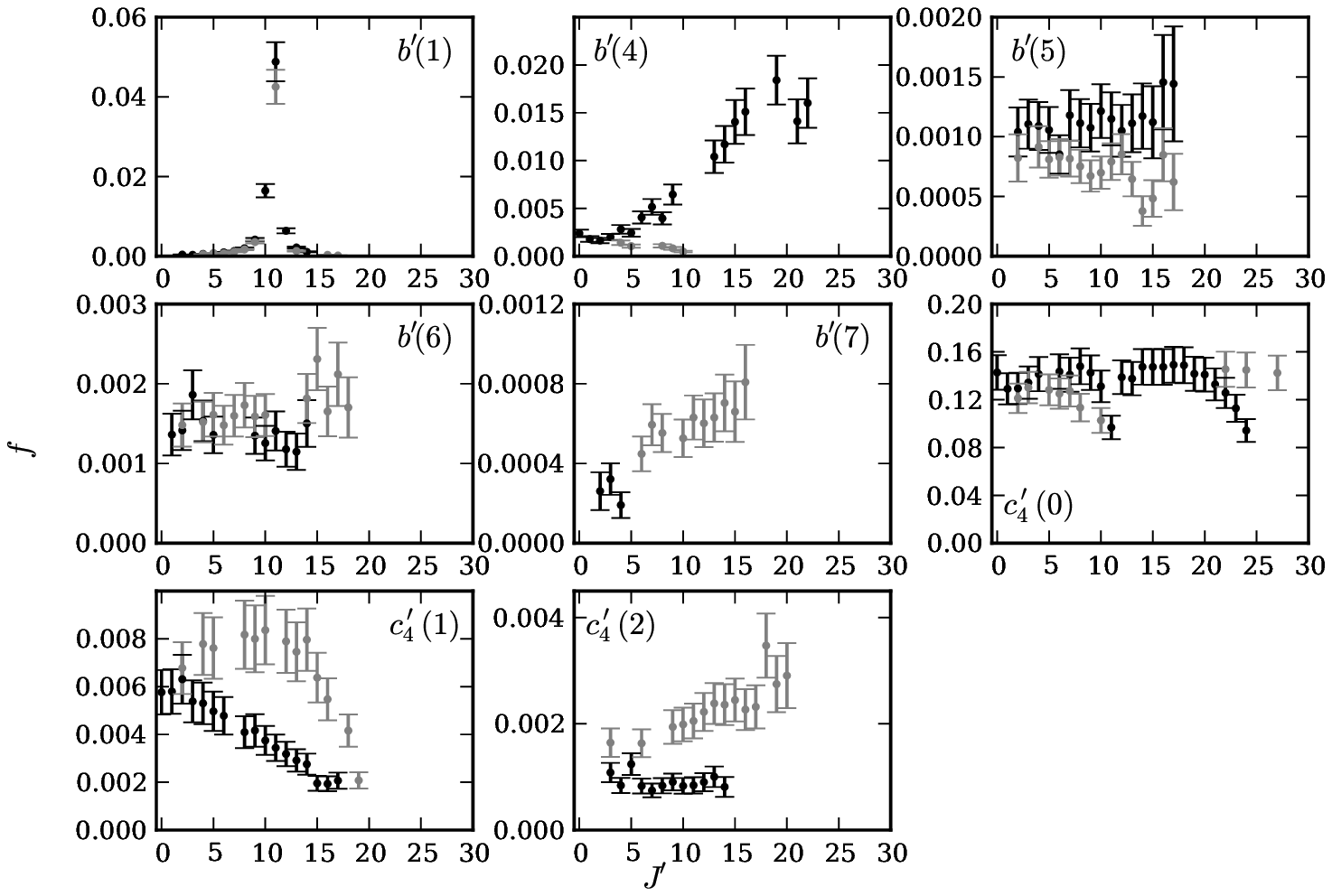}
  \caption{
    Rotational dependence of ${}^1\Sigma^+_u-X$ experimental ${}^{14}$N$^{15}$N $f$-values labelled according to their excited vibrational-level.
    Values are shown for $P$- and $R$-branch transitions \emph{(dark and light points, respectively)}.
}
  \label{fig:1Sigmau fJ}
\end{figure*}
Figure~\ref{fig:1Sigmau fJ} shows the rotational dependence of experimental $f$-values for the observed ${}^1\Sigma_u^+-X$ bands.
In all cases, there is a strong $J$-dependence of $f$-values and some degree of divergence of the two rotational branches.
Particularly large branching is seen for $b'(4)-X$, where the lack of measurements of $f$-values for $R$-branch lines with $J'>10$ is due to their complete disappearance from the spectrum.
Similarly, the large branching ratio for $b'(7)-X$ is not well highlighted in Fig.~\ref{fig:1Sigmau fJ} because of the completely annihilated $P$-branch for $J'>4$.

An example of strong state-interaction occurs for the group of bands $c_3(0)\,{}^1\Pi_u$, $c'_4(0)\,{}^1\Sigma^+_u$, $b'(1)\,{}^1\Sigma^+_u$, and $b(5)\,{}^1\Pi_u$ and has been analysed previously\cite{yoshino_tanaka1977,verma_jois1984,levelt_ubachs1992} with respect to ${}^{14}\rm{N}_2$ and ${}^{15}\rm{N}_2$.
A homogeneous interaction of $c'_4(0)$ and $b'(1)$ is evident in the present ${}^{14}$N$^{15}$N spectrum from a mutual repulsion of their deduced term series from a crossing point between $J=10$ and 11.
This perturbation is also evident in the $f$-values of Fig.~\ref{fig:1Sigmau fJ} where the very strong $c'_4(0)-X$ band is anomalously weakened near $J=10$, while the much weaker $b'(1)-X$ is greatly increased.
Without this interaction and the transfer of line strength from $c'_4(0)-X$ it is likely that $b'(1)-X$ would be too weak to be observed in the present spectrum, as $b'(0)-X$, $b'(2)-X$ and $b'(3)-X$ are.

There is also an evident $P$/$R$ branching of $c_4'(0)-X$ $f$-values, with the $P$ branch appearing stronger for $J<10$.
The reverse branching is evident in the $f$-values of $c_3(0)-X$, plotted in Fig.~\ref{fig:1Piu_fJ}, and these effects are broadly attributable to a heterogeneous interaction of $c'_4(0)\,{}^1\Sigma_u^+$ and $c_3(0)\,{}^1\Pi_u$ $e$-parity states so that the observed levels are of mixed ${}^1\Sigma_u^+$/${}^1\Pi_u$ character.
Then, the mixing of line strength factors for pure transitions of type $\Pi-\Sigma$ and $\Sigma-\Sigma$ leads to a constructive interference in the case of one rotational branch and destructive interference for the other.\cite{lefebvre-brion_field2004,gottscho_etal1978}
Furthermore, the highest-$J$ $c_4'(0)-X$ transitions in Fig.~\ref{fig:1Sigmau fJ} indicate a dominant $R$-branch.
This reversal with respect to low $J$ is not mirrored in the $f$-values of $c_3(0)-X$, indicating that $c_4'(0)$ is now principally perturbed by $b(5)$ which lies at higher energy and exhibits an appropriately strengthened $P$-branch.
The weakening of the $b(5)-X$ $R$-branch with respect to the $P$-branch for increasing rotation is dramatically evident in the experimental trace plotted in Fig.~\ref{fig:1415b05 example spectrum}.
The $Q$-branch of $b(5)-X$ is seen to have an intermediate strength because such ${}^1\Pi_u-X$ transitions terminating on $f$-parity levels are not prone to heterogeneous interaction with the ${}^1\Sigma_u^+$ states.

A similar degree of multi-state mixing is suggested by the $f$-values of $c'_4(1)$ and $b'(4)$ which may be principally attributed to their own homogeneous mixing as well as heterogeneous interactions with the nearby ${}^1\Pi_u$ levels $c_3(1)$ and $b(8)$.
Likewise, the observed heterogeneous perturbation of $b'(7)$ and $c'_4(2)$ $f$-values is primarily the result of rotational mixing with $b(11)$ and $c_3(2)$.

These examples illustrate the multi-level nature of perturbations that occur throughout the N$_2$ spectrum.
But because of the strength of the responsible electronic and rotational interactions, these are best explained in a global manner, such as by means of the CSE technique, rather than in terms of a local perturbation of a finite set of bands.

\section{Conclusions}

The results of a survey of ${}^{14}$N$^{15}$N XUV photoabsorption between 100\,000 and 109\,000\,cm$^{-1}$ are presented here and in an accompanying data archive\cite{heays_etal2011_data_archive} in terms of observed line-energies and linewidths, and deduced term-values and $f$-values.
All of these data were derived from measurements made with the FTS permanently fixed to the Desirs undulator source at the Soleil synchrotron, and this combination of an intense and broadband XUV source and very highly resolving spectrometer is well suited to the present study.
The revealed patterns of $f$-values and natural linewidths are complicated and perturbed at both the vibrational-band and rotational level, and differ markedly from observations of the other two isotopomers of N$_2$.
The effects of indirect predissociation are clearly evident in the measured linewidths, and the divergent $f$-values of rotational branches within the same band clearly indicate the rotational mixing of ${}^1\Pi_u$ and ${}^1\Sigma_u^+$ levels.

A CSE model was constructed without reference to the data collected here but its predictions are uniformly excellent when compared with the new measurements, providing further proof of the suitability of the technique for the modeling of the N$_2$ spectrum.
Work is under way to extend the CSE formulation to include $e$-parity excited states and a more complete interaction of ${}^1\Pi_u$ and ${}^1\Sigma_u^+$ states with the dissociative ${}^3\Pi_u$ states.
This work will be aided by observations in the present ${}^{14}$N$^{15}$N spectrum of previously-undetected direct perturbations between singlet and triplet levels, which are currently being analysed for a future report.

\section*{Acknowledgements}

This work has been carried out within the framework of the NWO astrochemistry program. The authors are grateful to the staff at Soleil for their hospitality and operation of the facility. 
The CSE calculations were supported by the Australian Research Council Discovery Program grants DP0558962 and DP0773050.


\end{document}